# We are IntechOpen,
# the world's leading publisher of Open Access books
# Built by scientists, for scientists

**5,100**
Open access books available

**126,000**
International authors and editors

**145M**
Downloads

**154**
Countries delivered to

Our authors are among the

**TOP 1%**
most cited scientists

**12.2%**
Contributors from top 500 universities

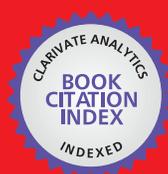

WEB OF SCIENCE™

Selection of our books indexed in the Book Citation Index
in Web of Science™ Core Collection (BKCI)

# Interested in publishing with us?
# Contact book.department@intechopen.com

Numbers displayed above are based on latest data collected.
For more information visit www.intechopen.com

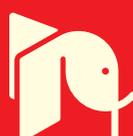

*Chapter*

# Interface Combinatorial Pulsed Laser Deposition to Enhance Heterostructures Functional Properties

*Jérôme Wolfman, Beatrice Negulescu, Antoine Ruyter, Ndioba Niang and Nazir Jaber*

**Abstract**

In this chapter we will describe a new development of combinatorial pulsed laser deposition (CPLD) which targets the exploration of interface libraries. The idea is to modulate continuously the composition of interfaces on a few atomic layers in order to alter their functional properties. This unique combinatorial synthesis of interfaces is possible due to very specific PLD characteristics. The first one is its well-known ability for complex oxide stoichiometry transfer from the target to the film. The second one is the layer by layer control of thin film growth at the atomic level using in-situ RHEED characterization. The third one relates to the directionality of the ablated plume which allows for selective area deposition on the substrate using a mobile shadow-mask. However PLD also has some limitations and important PLD aspects to be considered for reliable CPLD are reviewed. Multiple examples regarding the control of interface magnetism in magnetic tunnel junctions and energy band and Schottky barrier height tuning in ferroelectric tunable capacitors are presented.

**Keywords:** combinatorial synthesis, oxide thin films and multilayers, functional heterostructure

## 1. Introduction

The first report of laser ablation generated plasma to deposit a film dates back to 1965, using a continuous ruby laser [1]. As the obtained film quality was inferior to those made with other deposition techniques at that time, the laser deposition remained confidential for two decades. The discovery mid 80's of the high Tc superconductor $YBa_2Cu_3O_{7-x}$ (YBCO) [2–3] triggered a frantic search for room temperature superconducting cuprates materials, including strong effort for thin film synthesis. The complex cuprate cationic composition makes film growth difficult using conventional physical vapor deposition (PVD) techniques like magnetron sputtering or evaporation. Indeed, to grow films with the right composition it is mandatory to start with a stoichiometric plasma. The $Ar^+$ sputtering rate of multi-cationic targets and the distribution in the plasma strongly depends on the cations mass, which precludes the use of a stoichiometric target to obtain a stoichiometric plasma. The offset target composition leading to stoichiometric films is unpredictable and a trial-and-error method is usually employed that renders magnetron sputtering an impractical technique for multi-cationic new materials research. Evaporation temperatures depend on the nature of the cations, which make a thermal vaporization of stoichiometric target inappropriate too. The attempt to produce cuprates films using sputtering and evaporation led to poor quality films.

Ceramic target vaporization based on laser ablation does not depend on the nature of the atoms within the target and result in a stoichiometric plasma as long as the energy per surface unit (i.e. the laser fluence) is above the ablation threshold. Venkatesan and co-workers were the first to vaporize an YBCO target using a pulsed excimer laser (UV KrF). After fine tuning the ablation conditions by checking the spatial distribution of the cations they rapidly produced superconducting films having the best physical properties of that time. [4] This first successful synthesis of high Tc cuprates films using a pulsed laser drew the interest of the community and several laboratories started working in the field now known as pulsed laser deposition (PLD). In the following decade, beside cuprates superconductors, strong interest in colossal magneto-resistive manganites and multiferroic ferrites has arisen. This intense scientific activity on multi-cation oxides associated to PLD ease of use, versatility and reasonable cost contributed to its rapid expansion in the 90's.

Since then PLD based thin film research thrived over a wide range of materials, physical properties and applications. PLD has benefited from in-situ real-time characterization tools developed for other deposition technique to mature into an epitaxial film growth method with control at the atomic scale and able to produce heterostructures with sharp interfaces. At the end of the 90's, material scientists considered transposing combinatorial synthesis, a high throughput synthesis method developed by the pharmaceutical industry, to the field of multi-cation oxides research. Combinatorial PLD (CPLD) rapidly emerged, the basic idea being to grow on a single substrate, in a single deposition run, a film with continuous in-plane chemical modulation. In this way, a wide range of chemical compositions are produced within the same sample and can be quickly scanned to identify compounds with optimum targeted properties. Since the new millennium, CPLD has been refined and its field extended to ternary phase diagram exploration. More recently CPLD research field has been extended to a new territory: the exploration of interface compositions in heterostructures with enhanced functional properties a.k.a. Interface Combinatorial Pulsed Laser Deposition (ICPLD).

## 2. Pulsed laser deposition for combinatorial synthesis

The aim of this chapter is not to thoroughly describe PLD but to emphasize its most important aspects, advantages and limitations with regard to combinatorial synthesis of oxide films and heterostructures. The reader interested in an exhaustive description of PLD is encouraged to consult PLD's introduction reference book [5] or the other chapters devoted to the subject in this book.

The first pre-requisite for CPLD is the growth of films with uniform thickness and homogeneous composition over the entire sample surface. Although PLD has a reputation for stoichiometric transfer from the target to the film, this is not however straightforward and several deposition parameters have to be fine-tuned. Starting from the formation of a stoichiometric plasma, target thermal vaporization occurring below the ablation threshold should be reduced as much as possible. This means that local fluence everywhere on the laser beam spot should be above the ablation threshold, implying very steep sidewalls of the laser beam energy distribution. Such a distribution, called a top-hat, necessitate laser beam shaping with a beam homogenizer. The laser beam divergence depends on the discharge voltage and could affect the beam spot size and energy distribution depending on the beam shaping method, so the discharge voltage should be kept constant. The plasma expands from the target toward the substrate in the form of a plume which interacts with ambient gas molecules or atoms. Multiple collisions per atom or ions occur leading to a thermalized but still highly directional plasma reaching the substrate. It follows a radial distribution of thickness and composition at the substrate surface. In order to get a uniform and homogeneous film, one has to scan the plume with respect to the substrate, by either scanning the laser spot over the target surface at constant focus and/or moving the substrate (rotation, translation). Doing so will result in a film with much increased thickness uniformity and homogeneous composition. However the composition may not match the target stoichiometry. Indeed, it has been shown that cationic off-stoichiometry up to a few percent occurs depending on the fluence for a fixed spot size and target-substrate distance [6]. So cationic composition/uniformity should always be checked versus fluence with high sensitivity characterization techniques. Once the right fluence has been determined and stabilized over the laser spot on the target, it is crucial to ensure that it stays constant in time. Regarding this point, two factors have to be taken into account. The first one is the degradation of the excimer laser gas charge with time, leading to a drop of the output energy. As already specified, it is advisable to work at constant voltage discharge, so one should not compensate the energy drop with voltage. The solution is to start with a higher energy output than required and to modulate the beam energy with an external attenuator. The second factor is the laser attenuation at the chamber introduction viewport. Although the visible part of the plasma, i.e. the plume, is very forward peaked, some low energetic species are slowly deposited everywhere in the chamber, including on the introduction viewport. This leads to a time varying attenuation which will affect the fluence at the target surface. To compensate for this attenuation, it is mandatory to measure the fluence after the viewport and to increase the input energy accordingly. One can reorient the beam out of the chamber through a pollution preserved port with a translatable UV mirror to measure its energy without breaking the vacuum.

The second pre-requisite for CPLD is a layer by layer growth mode for a control of the composition at the unit cell level. Layer by layer PLD growth of various perovskites has been demonstrated multiple times using in-situ real time grazing incidence electron diffraction (RHEED). This ability for 2d growth comes from the very peculiar and sequential PLD surface crystallization kinetic. Indeed, although average PLD deposition rate is quite slow, instantaneous deposition rate a few μs after laser pulses is extremely high, creating a supersaturation close to the surface which results in high nucleation rate. The large number of nucleus present after the first pulse favors 2d growth. Furthermore, as PLD relies on photon to vaporize the target, the ambient gas pressure can be varied over a very wide range from vacuum level to a few mbar. The deposition pressure allows for the control of the energy of the species reaching the substrate and ultimately their remaining kinetic energy to explore the surface and find the nucleation sites.

The last pre-requisite for CPLD is the ability to lower the average deposition rate per pulse to produce the smallest composition step when mixing materials from different targets. For a fixed fluence and target-substrate distance, the deposition rate per pulse strongly correlates with the laser spot size on the target which is easily adjustable. Deposition rate as low as a few hundred pulses per perovskite unit cell thick layer can be reached. One drawback of PLD is the lack of deposition rate stability. Indeed the structure of the irradiated target surface evolves with laser exposure; the deposition rate lowering with target aging is more severe for a brand new surface and decays with target exposure. Scanning over a large area of the target reduces the number of laser spots per location and makes this problem less stringent. It is however mandatory to calibrate the deposition rate prior a deposition or to measure it in-situ using RHEED.

## 3. Combinatorial pulsed laser deposition

The concept of combinatorial research, initially introduced in the pharmaceutical industry, aims at synthesizing a large number of compounds with complex and systematically varied composition in a single batch. A high speed characterization technique is then used to scan through this material library and identify the compounds presenting the targeted property.

This approach is very appealing regarding multi-cationic perovskite-related oxides. Indeed their physical properties, covering a very wide span (superconductivity, ferroelectricity, ferromagnetism, colossal magnetoresistance, tunable resistivity …), result from the subtle equilibrium between competing interactions involving charges, spins and orbitals. This results in an extreme sensitivity of the properties to the cationic composition and requires a fine, thorough and systematic scanning in order to optimize the performances.

The first thin film combinatorial synthesis attempts implied multiple successive room temperature depositions to vary the composition. Several annealing steps were then required for thermal diffusion and crystallization. This approach, where reaction products are dominated by thermodynamic like in ceramic sintering, is not appropriate for epitaxial thin films and heterostructures. On the contrary, epitaxial oxide thin films growth by PLD does not require post-deposition annealing. Moreover the directionality of the plume allows for localized deposition on the substrate through a shadow-mask. These multiple advantages were soon associated to combinatorial synthesis of oxide materials at the end of the 90's simultaneously in the United States and in Japan. The original approach is schematically described in **Figure 1**.

The single unit cell combinatorial cycle consists of two successive depositions from different targets. The material deposited from each target is distributed across the substrate surface using a moving shadow-mask in order to vary locally the layer completeness (from 0–100%). After one deposition cycle, the resulting 1uc thick layer has a composition which varies laterally from one target composition to the other (composition spread). Then this deposition cycle is repeated N times in order to achieve targeted final film thickness.

This idealized combinatorial PLD synthesis produces a continuous variation of compositions on a single sample, guaranteeing identical growth conditions for all compositions and eliminating the risk of sample variability.

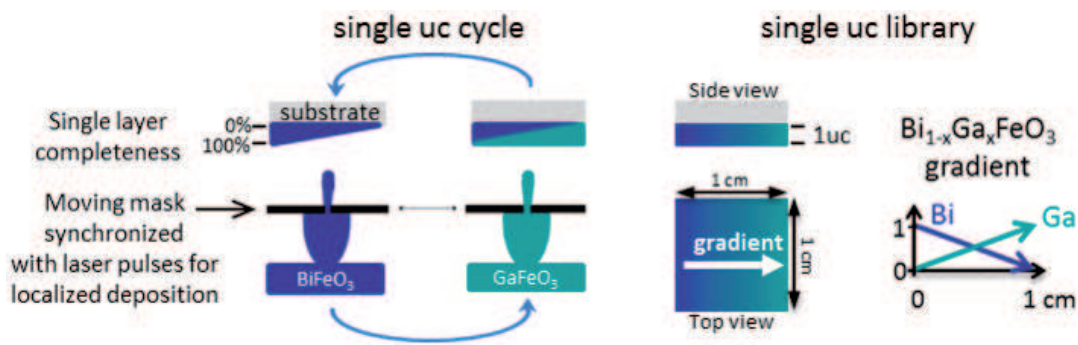

**Figure 1.**
*Schematic of the combinatorial single layer synthesis and of the resulting composition spread library (e.g. $Bi_{1-x}Ga_xFeO_3$ composition library).*

A consequence of CPLD is the necessity to use local probes to scan through the compositions library and assess performances. There is a trade-off between the minimum probe size required to measure the targeted physical property and the lateral composition gradient in order to characterize a "homogeneous" compound at the probe level.

The main difficulty regarding CPLD synthesis is to reach adequate control of both local composition and thickness using PLD. As explained in the previous chapter this requires long in-depth preparation work involving chemical characterization and exhaustive optimization of all the deposition parameters.

A prerequisite to reliable CPLD synthesis is to produce a uniform and smooth film with constant composition and thickness from each target over the surface of the future CPLD samples. One should not attempt CPLD before proving this achievement. Another important point is to keep a statistical approach to the characterization of the libraries. It is tempting to produce ternary phase diagrams using 3 different targets on a single substrate. However, in this case there is only one location on the sample per compound which is statistically insufficient. To our knowledge, any published work concerning CPLD ternary phase diagrams on a single sample does not present a proper characterization that demonstrates the control of the composition and thickness across the library. This is not very surprising knowing how difficult it already is to master a single gradient for a binary diagram across one sample. Unfortunately even for binary samples a large number of CPLD articles have been published without any evidence of control of composition/thickness which brought some discredit on CPLD among the scientific community. To change this perception and reinforce CPLD credit, we proposed a decade ago a statistical approach to the characterization of both the composition and the physical properties for binary phase diagram. [7] More recently we developed an alternative approach to explore ternary phase diagrams: instead of trying to produce the full ternary phase diagram on a single sample, we select lines of compositions cutting through the corresponding triangular diagram. Each synthesized sample has a composition gradient in one direction with the compositions defined for one line. [8] This way statistical characterization are possible along the direction orthogonal to the gradient. Scanning through the ternary phase diagram along multiple lines thus requires the synthesis of a few samples with only three targets.

To illustrate the effectiveness of CPLD we will discuss about the search for new lead-free piezoelectrics to replace $(1-x)PbZrO_3-xPbTiO_3$ (PZT), the most used material in microelectromechanical systems for sensing, actuating, or energy harvesting applications. PZT ferroelectric films present large piezoelectric coefficients and electromechanical coupling, enabling long range motions and high energy densities. [9] A unique characteristic of lead-based solid solutions presenting high piezoelectric coefficients is the strong enhancement of their piezoelectric response in the vicinity of a composition induced phase transition between ferroelectric phases with different crystalline symmetries, called a morphotropic phase boundary (MPB). In PZT the MPB lies between a rhombohedral ferroelectric phase and a tetragonal ferroelectric phase. [10] The microscopic origin of this enhanced piezoelectric activity is still being debated but usually involves easiness of polarization rotation at the MPBs. [11–13] PZT is lead-based and thus targeted by environmental regulations (e.g. RoHS EU Directive). Thus, alternative piezoelectric lead-free materials are required, and an obvious direction is to look for MPBs in other ferroelectric solid solutions. The rhombohedral perovskite $BiFeO_3$ (BFO), being a robust ferroelectric (Tc ~ 1100 K) with record polarization (100 μC·cm$^{-2}$), is a good starting point. [14, 15] Solid solutions of BFO with tetragonal ferroelectric perovskite phase like $PbTiO_3$ and $BaTiO_3$ have been synthesized, and MPBs have been found in both cases. [16, 17].

As previously discussed, the first step is to produce uniform films of controlled composition before aiming to CPLD. The Bi element being volatile, the stabilization of pure BFO in thin films is not straightforward, as several parasitic phases can coexist. To compensate for Bi volatility we used an enriched $Bi_{1.1}FeO_3$ target and found the range of temperature, oxygen pressure and fluence which lead to pure BFO films. The structure was studied by x-ray micro-diffraction (µXRD) and Bi/Fe ratio by Rutherford Back Scattering (RBS) versus laser fluence. A fluence of 1.72 J·cm$^{-2}$ was identified as giving Bi/Fe = 1 with a deposition temperature of 700°C, an oxygen pressure of 0.2 mbar, a laser repetition rate of 6 Hz with a target-substrate distance of 4.5 cm. [18] RBS is an averaging technique (spot size of 2 x 2 mm$^2$), so to assess local uniformity of thickness and composition we turned to an Electron Probe Micro-Analyzer (EPMA) equipped for Wavelength Dispersive X-Ray Spectroscopy (WDS) (see [18] for details). A specific thin film analysis program has been used to determine BFO's composition (TFA/WDS layerf, Cameca). BFO film's density × thickness product (ρ·t) and composition were simultaneously computed. Thirty measurements were realized with a beam diameter of 20 µm (20 keV, 100 nA) every 300 nm along the film. **Figure 2a** represents the ρ.t product and the weight percentages of Bi, Fe and O after self-consistent analysis of the raw data. The average value of ρ.t is 242 µg/cm$^2$ with a standard deviation of 5.4 µg/cm$^2$ equivalent to a relative variation of 2.2%. Considering the bulk BFO density (d = 8.38 g/cm$^3$) we find a thickness t = 289 nm ± 6.5 nm (1σ). The stability of composition is even greater along the sample. The average weight percentages and standard deviations for the Bi and Fe, transposed into atomic percentages are respectively 20.02% ± 0.08% and 19.98% ± 0.08% for an expected value of 20%. With these statistical analysis, we find that the composition dispersion is Bi$_{1.001 \pm 0.004}$Fe$_{0.999 \pm 0.004}$O$_3$ (1σ) and thickness standard deviation σ(t) ≤ 2.2% i.e. a very good thickness and composition uniformity along the sample surface. As the aim was to measure the piezoelectric coefficient of BFO-based solid solution, we deposited BFO onto La$_{0.8}$Sr$_{0.2}$MnO$_3$ (LSMO), an epitaxial conductive oxide electrode grown on SrTiO$_3$ (001) substrate. A zoom of the θ-2θ X-ray diffraction pattern around (001) diffraction peak is shown in **Figure 2b**. Only (0 0 l)$_{pc}$ (pseudo-cubic notation) oriented diffraction peaks are visible on this pattern while no parasitic phase could be detected [19]. The thickness (Pendellösung) fringes observed around both LSMO and BFO (001) reflections demonstrate the crystalline quality and the smoothness of surface and interfaces.

We choose GaFeO$_3$ (GFO) as the second member of the solid solution to be explored in order to find a MPB. GFO does not have a perovskite structure but crystallizes in a much more complex orthorhombic structure (SG Pc21n). [20] In the (1-x)BFO – (x)GFO system (BGFO) designed here, $Ga^{3+}$ cations have incentives to substitute for $Bi^{3+}$. Small ions like $Ga^{3+}$ do not occupy perovskite A-site in conventionally synthesized ceramics. However, perovskites with small A-site cations like $Sc^{3+}$ and $Mg^{2+}$ have recently been stabilized using high-pressure high-temperature synthesis, and $Ga^{3+}$ is envisaged as A-site cation in this emerging field. [21] Epitaxial strain during PLD growth has longtime proven to be an alternative to high pressure synthesis for metastable phase stabilization. [22] As GFO is not a perovskite, we do not expect to obtain a solid solution at high x values, so we limited our range of investigation to $0 \leq x \leq 0.12$. In this range, the Goldsmith tolerance factor t is greater than 0.87, not too far from non-substituted BFO's tolerance factor (t = 0.89), and compatible with a distorted perovskite structure. So, it is plausible that some $Ga^{3+}$ ions occupy the perovskite A-site in our films, although it is probable too that part of the $Ga^{3+}$ shares the B-site together with $Fe^{3+}$. The $Ga^{3+}$ substitution for $Bi^{3+}$ being limited to 0%-12%, the BFO deposition conditions were used for GFO. WDS analysis were done along and across the composition gradient (1 point each 300 μm). Assuming the $Bi_wGa_xFe_yO_z$ formula with z = 3, cationic contents w, x, and y have been extracted from these measurements, the error being estimated to 0.005. The extracted Ga content is plotted versus position in **Figure 3a**, showing a linear increase from 0% up to 12%, in good agreement with the nominal concentrations. We note that both Bi and Fe contents decrease from 1 in pure $BiFeO_3$ to about 0.97 for 6% Ga doping. As x increases from 6–12%, Fe content goes back to about 1.0 while Bi content continue to decrease. [19] From these values, one could suspect that Ga is substituted partly for Bi and partly for Fe.

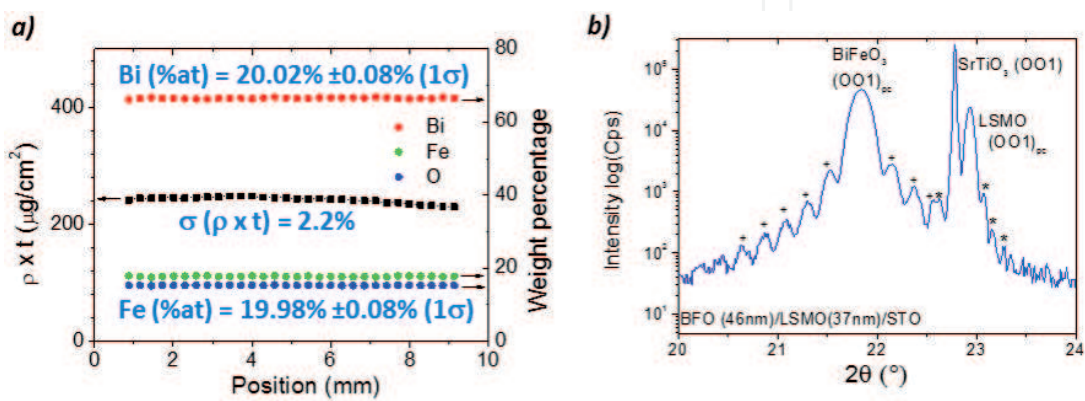

**Figure 2.**
*(a) Evolutions of density × thickness product (ρ·t) and weight percentages of Bi, Fe and O as a function of the position (adapted from [18] with permission) and (b) X-ray diffraction pattern of BFO/LSMO/STO heterostructure.*

X-ray Reciprocal Space Mapping (RSM) around the $(103)_{pc}$ reflection have shown that BGFO and LSMO are epitaxial on STO (not shown here, see [19]). The lattice parameters evolutions confirm that Ga effectively enters into the BFO structure. Furthermore a characteristic splitting of the $(103)_{pc}$ reflection of BFO strained by cubic STO disappears for a Ga content 5% < x < 7%. This indicates a change of symmetry of the film and might be the signature of a MPB. Piezoelectric characterizations were made on 30 x 30 $\mu m^2$ top Pt electrodes (dc-sputter deposited via a lift-off process) using a laser scanning vibrometer (LSV model MSA-500, Polytec, $V_{AC}$ = 1 V) (see schematic of the heterostructure **Figure 3b** top). A typical mapping of the extracted effective piezoelectric coefficient $d_{33}^{eff}$ across one electrode, using a 3 μm laser spot size, is presented in Figure 3b (bottom), showing a uniform displacement. The $d_{33}^{eff}$ coefficients were extracted from fifteen LSV measurements on each electrode and three different lines across the composition gradient were measured. **Figure 3c** shows the variation of the $d_{33}^{eff}$ as a function of Ga content with standard deviation represented as error bars. After a slow increase of $d_{33}^{eff}$ at low Ga content, a sharp peak centered at about 6.5% is observed. The maximum $d_{33}^{eff}$ value, about 53 pm/V, is twice larger than the value obtained for undoped BFO. Finally, above x = 8%, $d_{33}^{eff}$ falls to 15 pm/V. The sharp enhancement of piezoelectric properties occurring around 6.5% of Ga doping is correlated with the symmetry change observed around the same composition by RSM. [19] To confirm the presence of a MPB we looked for a change in the ferroelectric properties of BGFO with Ga content x. We used a ferroelectric tester (Radiant LC II) to measure polarization hysteresis P(E). The hysteresis cycles presented **Figure 4** were acquired at liquid nitrogen temperature.

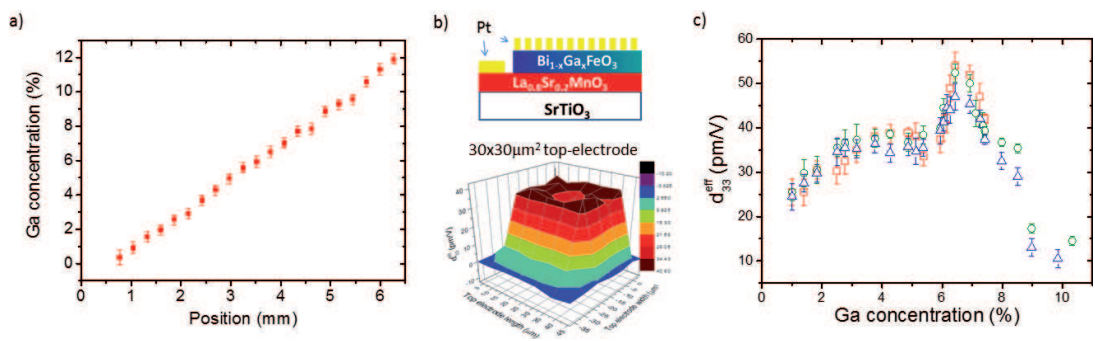

**Figure 3.**
*(a) Ga concentration measured by WDS along the composition gradient. (b) Heterostructure schematic (top) and $d_{33}^{eff}$ mapping of a top electrode measured by laser vibrometry (bottom). (c) BGFO $d_{33}^{eff}$ as a function of Ga content. From [18] with permission.*

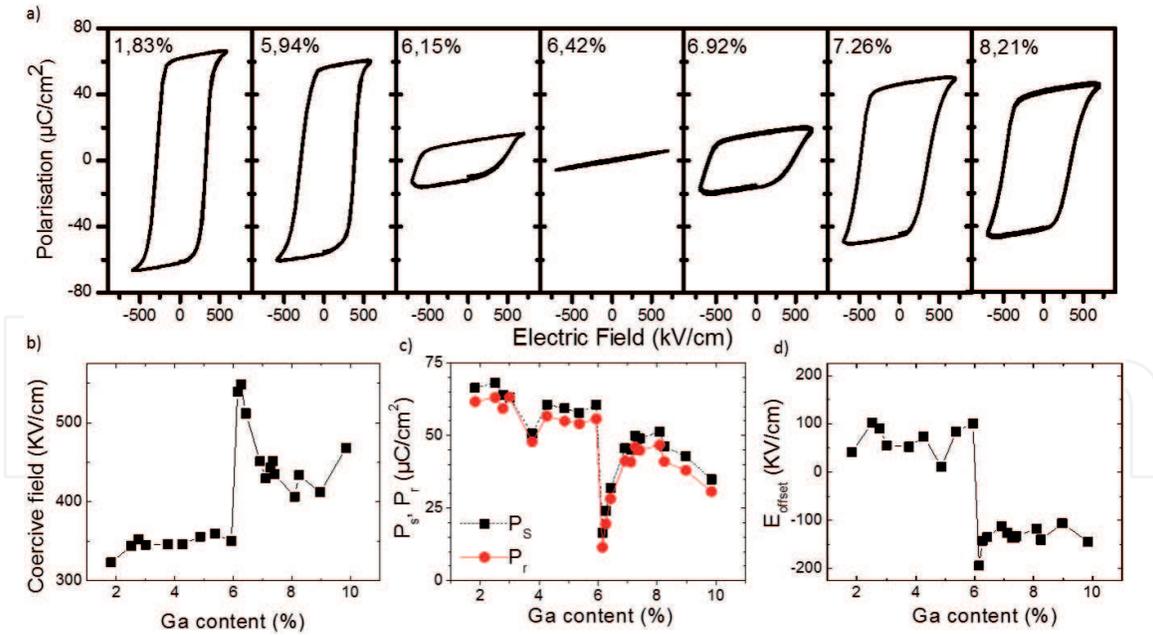

**Figure 4.**
*(a) Ferroelectric cycles for various Ga contents measured at 77 K. the corresponding coercive fields, remanent and saturation polarizations and electric field offsets are plotted in (b), (c) and (d) respectively.*

A clear transition in the ferroelectric cycle shapes is visible **Figure 4a** as Ga content increases. As the Ga content gets over 6%, a strong increase of the coercive field (**Figure 4b**) associated to a strong decrease of both saturation and remnant polarization (**Figure 4c**) are observed, together with a change of sign of the electric field offset (hysteresis imprint **Figure 4d**). This demonstrates that a change of ferroelectric phase is occurring, correlated to the $d_{33}^{eff}$ peak, and implies that a MPB is present in BGFO at about 6.4%. It is important to note that the $d_{33}^{eff}$ peak is very sharp in composition and that a conventional ceramic approach with 1% doping steps would have miss it, emphasizing the power of the continuous composition spread in CPLD.

## 4. Interface combinatorial pulsed laser deposition

### 4.1 Interfaces of oxide heterostructures: the new territory

In strongly correlated complex oxides, charge, spin, orbital and lattice degrees of freedom co-exist and interplay cooperatively. In particular the complex balance between these degrees of freedom and related interactions generates a rich spectrum of competing phases in perovskites or perovskite-derived materials (e.g. high Tc superconductor, metal–insulator transitions, magnetism, ferroelectricity, piezo-electricity…). The recent progress in deposition techniques allowed the production of complex perovskites heterostructures with atomically sharp interfaces, which expanded material researcher's horizon. Fascinating phenomena and novel states of matter at complex oxide heterointerfaces have been reported. One can cite for instance the existence of high mobility 2d electron gas at $LaAlO_3$ / $SrTiO_3$ interface, even becoming superconducting at low temperature, while both materials taken separately are insulating. [23, 24] Another striking example is the transition of $CaTiO_3$ from its usual non-polar state into a high-temperature polar oxide thanks to interfacial tilt epitaxy [25]. The isomorphism of the $ABO_3$ perovskite oxide structure allows for a wide range of chemically modulated interfaces.

Some of the phenomenon occurring at perovskite interfaces are reported **Figure 5**. Rumpling, polar discontinuity, interfacial B-site cation environment asymmetry, $BO_6$ octahedral rotations are all potential levers to modulate interface properties. Their complex interplay is strongly affected by cationic substitutions and a complete and fine exploration of the possible interface compositions is required in order to identify new physics phenomena or enhanced properties. Interface CPLD (i.e. ICPLD) is a powerful tool in that respect.

One application where oxide interfaces plays a crucial role is the ferroelectric (FE) voltage tunable capacitor envisaged for future RF communication technologies (5G and Near Field Communication NFC). [27] The relative dielectric permittivity $\varepsilon_r(E)$ of a FE has a large electric field dependence. [27] The perovskite solid solution $Ba_{1-x}Sr_xTiO_3$ (BST) is the most widely used FE in current 4G thin film parallel plates varactors because of its excellent tunability/losses compromise. New specifications for 5G and NFC (higher frequency and reduced driving voltage) call for improved varactor properties. Reducing the FE film thickness from 240 nm (4G) down to 50-100 nm range is one option toward meeting these new specifications. However, in such thin FE films the metal / FE (M/FE) interface influence is reinforced in a damaging way, due to two interfacial phenomena. The first one is the existence of FE "dead-layers" with degraded $\varepsilon_r$ and spontaneous polarization close to the electrodes, producing an effective non-tunable interfacial capacitance. [28, 29] The second one is the increased leakage current due to insufficient Schottky Barrier Height (SBH).

Interface engineering can be used to tailor band alignment and interface polarizability. The insertion of a thin layer with different atomic element(s) at the interface allows to manipulate the chemical bonding and promotes atomic rearrangement. Let us consider for instance the anti-displacement of anions and cations predicted at $Ba^{2+}O^{2-}$/M and $Sr^{2+}O^{2-}$/M interfaces and quantified by a rumpling parameter R. [30] R depends on the chemical bonding and is responsible for an interface dipole, which in turn modulate the SBH. Interestingly, the insertion of e.g. a single Al atomic plane at the BaO/M interface strongly affects R and SBH. Indeed, for M = Pd the SBH goes from 1.4 eV to 2.6 eV. [30] Significant rumpling has been experimentally shown for $SrTiO_3$ (STO) in contact with $La_{2/3}Sr_{1/3}MnO_3$ (LSMO), a metallic perovskite electrode, inducing a polarization in the non-ferroelectric STO. [31] The continuity of the perovskite structure through the LSMO/STO interface and its ionic character offer new ways to control electronic properties. In $La_{1-x}Sr_xMnO_3$ ($LSMO_x$), the B-site cation ratio $Mn^{3+}/Mn^{4+}$ is determined by the A-site ratio $La^{3+}/Sr^{2+}$. Along [100], successive AO and $BO_2$ planes are polar for LSMOx and charge neutral for BST. Interfacing LSMOx with BST leads to tunable interfacial polar discontinuity which can induce lattice polar distortion and result in SBH modulation. [32–34].

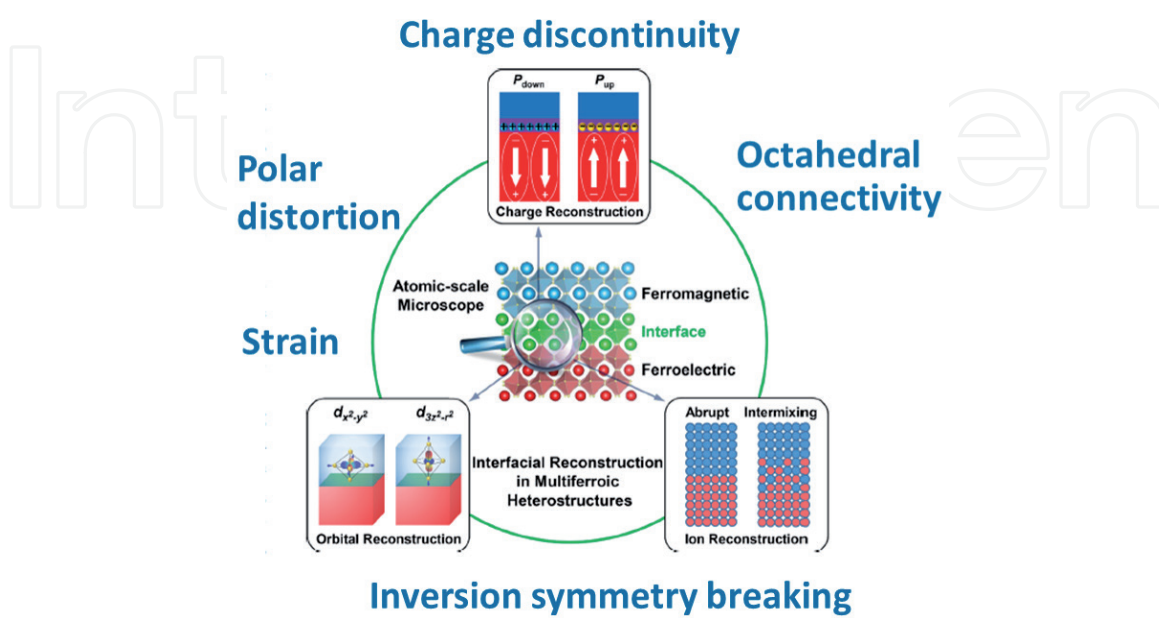

**Figure 5.**
*Schematic of competing interactions and phenomenon at perovskite $ABO_3$ interfaces. From [26] with permission.*

LSMO is a ferromagnetic (FM) half-metal, i.e. having a 100% spin-polarization at the Fermi level. For the latter reason it has been intensively studied as a spin-polarized electrode in LSMO/STO/LSMO magnetic tunnel junction (MTJ). MTJs are used e.g. as memory bits in magnetic MRAMs. The tunnel resistance depends on electrode spin-polarization and on the relative orientation of the electrode magnetic moments, with high resistance $R_{AP}$ (resp. low resistance $R_P$) for antiparallel (resp. parallel) states. A 100% spin polarized electrode leads to a theoretical infinite $R_{AP}$ which is ideal for the cited application. In LSMO/STO/LSMO, a record tunnel magneto-resistance (TMR = $(R_{AP}-R_P)/R_P$) of about 2000% was reported, but unfortunately for temperature far below the Curie temperature $T_C$. [35] The vast majority of the electrons tunnel from the interfaces, their spin-polarization being affected by the nature of the chemical bonding. FM correlations at manganite interfaces are known to be weaker than in bulk, causing a magnetic "dead layer" which probably explains the diminution of TMR close to $T_C$. [36–38] Attempts have been reported at creating a doping profile at the interfaces by inserting a 2 uc thick $LaMnO_3$ layer [39, 40] or a single uc thick $La_{0.33}Sr_{0.67}MnO_3$ [41] layer to overcome this problem with some improvement of interface magnetism but still not a full recovery of bulk properties. As for SBH and interface polarizability, multiple factors might participate to interface magnetism weakening, like charge discontinuity driven intermixing, octahedral tilt induced in the first LSMO layers by octahedral connectivity at the interface, substrate strain and so on. A combinatorial heuristic approach to the definition of interface composition is a powerful tool to help understanding all these factors interplay and to enhance the interface magnetism, SBH or interface polarization.

**4.2 The LSMO/STO interfaces**

Incorporating a few uc of combinatorial $LSMO_x$ ($0 \leq x \leq 1$) at the LSMO/STO interface to modulate the chemical bonding, the carrier density and the polar discontinuity could potentially induce STO lattice polar distortion, SBH modulation, as well as restoring interface ferromagnetism.

*4.2.1 Ferromagnetism at STO/LSMO interface*

Before producing the described ICPLD heterostructures, we first optimized the LSMO physical properties, composition and thickness uniformity. The magnetic properties of the film were then characterized versus temperature using a commercial Kerr magnetometer equipped with a cryostat (NanoMoke II, Durham Magneto Optics). As the magnetism at the LSMO/STO interface is weakened, the Curie temperature will depend on LSMO thickness for very thin films. To avoid this regime, we worked with 30 nm thick LSMO films (~80uc). Several films were deposited on $TiO_2$ terminated (100) STO substrates with high-pressure RHEED monitoring (Staib/TSST) at various fluence, temperature and oxygen pressure. The optimized deposition conditions leading to a $T_C$ = 341 K were $P_{O2}$ = 0.2 mbar, $T_{sub}$ = 850°C, f = 5 Hz and a fluence of 0.83 J/cm$^2$. RHEED oscillations were clearly visible during all the deposition process implying a layer by layer growth. X-ray diffraction patterns (Θ-2Θ) showed only (0 0 l)$_{pc}$ reflections with thickness fringes attesting for the crystalline quality and the surface and interface smoothness. RSM confirmed epitaxial "cube on cube" growth of LSMO on STO. The homogeneity of the films in term of composition and magnetic properties over a 1 cm$^2$ STO substrate was verified for thinner films, in the range where $T_C$ is thickness dependent. A 20uc thick sample was deposited with vertical and horizontal scanning of the laser, staying in focus at the target position, and of the substrate respectively.

The Kerr magnetometer laser spot (diameter < 5 μm) was scanned on the sample surface at fixed temperatures to measure magnetic hysteresis curves in 311 points spread across the sample surface. This (x,y) scan was repeated every 2.5 K from room temperature to 350 K after thermal stabilization. Each hysteresis curve was processed in order to extract saturation and remnant magnetization ($M_{sat}$ and $M_r$ respectively). M(T) curves can then be reconstructed for each point on the sample surface, allowing to assess for Tc in each location. Maps reporting FM and paramagnetic (PM) areas of the sample are reproduced in **Figure 6a** (top) for various temperatures, the measurement points being indicated with black dots. The distribution of Tc is reported as a FM area percentage in **Figure 6a** bottom. Over 91% of the surface transit from FM to PM states on a temperature range less than 5 K wide (325 K < T < 330 K) and 100% inside a 10 K range. As LSMO's $T_C$ is very thickness and composition sensitive, the tight $T_C$ distribution indicates a good composition and thickness uniformity. We confirmed this uniformity with a WDS characterization over a 9x9 mm$^2$ area (25 x 25 = 625 points) of the same sample for La, Sr and Mn (JEOL 8530F). The small film thickness conjugated to the presence of Sr in the substrate did not allow to compute the composition with cationic ratios of the film. However, the WDS sensitivity is high enough to provide maps of relative variations for each element (see **Figure 6b**). The Sr map, with signal originating mostly from STO substrate, illustrates the electron beam stability ($\sigma_{Sr}$ ~ 1%) which is crucial for point to point comparison. Note that the drop in the corner, corresponding to silver paste contact to evacuate the charges, was excluded from the statistical analysis. On La and Mn maps a similar slight slope is visible with a corresponding standard deviation of 4.6% and 4.9% respectively. WDS signal is strongly correlated to the thickness, therefore we can conclude that the thickness distribution is relatively tight.

The relative interface contribution to the overall magnetic signal increases as the LSMO thickness decreases. It is however difficult to predict the optimal LSMO thickness leading to an improved interface contribution detection as the overall magnetic signal also decreases with thickness. A powerful aspect of CPLD is the possibility to deposit wedge-shaped layers with continuous thickness variation using shadow-masking (see **Figure 7a**). Before inserting the ICPLD layer, we checked the thickness control on two LSMO wedges, spanning from 8 uc to 76 uc, by measuring $T_C$ versus (x,y) and temperature. The obtained $T_C$ are represented **Figure 7b** with standard deviation represented as error bars. Tc noticeably decreases below 30 uc with an acceleration below 20 uc. In the inset of **Figure 7b** is represented a $T_C$ map of wedge#2, with the measured points represented as black dots. Constant nominal thickness levels are vertical, with thickness variation along x. A 10 K color increment is used and one can see that the lines separating the adjacent areas are almost vertical, attesting the good control of thickness variation in the wedge.

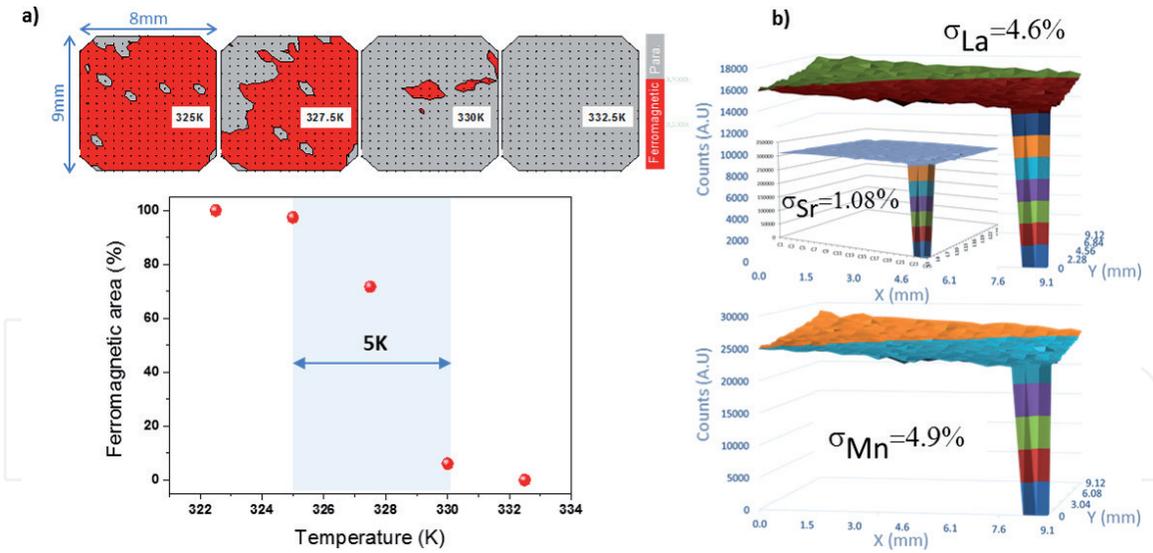

**Figure 6.**
*(a) Magnetic state maps of LSMO 20 uc film at various temperature (top) and $T_C$ distribution at the sample surface (bottom). (b) WDS signal for La, Sr and Mn over 9x9 mm² of the same sample.*

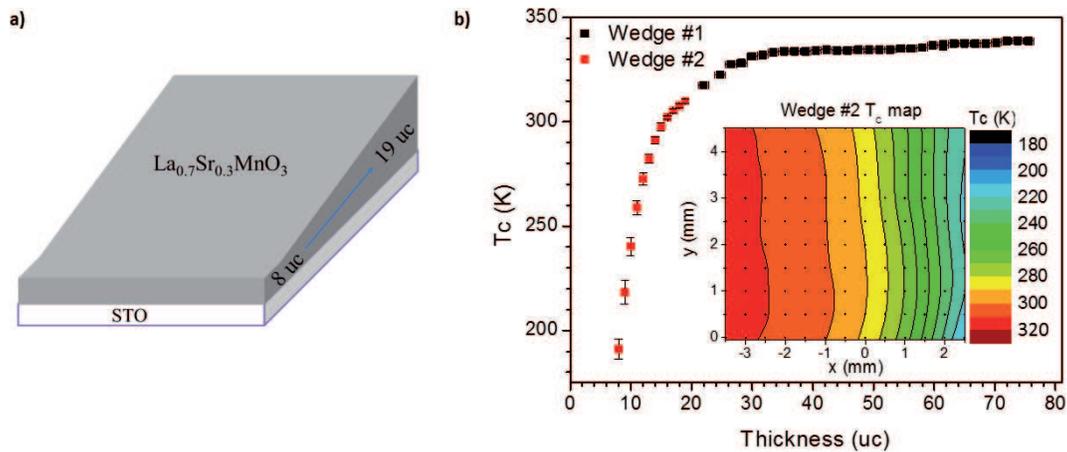

**Figure 7.**
*(a) Schematic of LSMO wedge#2. (b) $T_C$ versus thickness and (inset) $T_C$ map for wedge #2.*

To synthesize the ICPLD LSMOx layer we used $LaMnO_3$ (LMO) and $SrMnO_3$ (SMO) targets with the deposition parameters identified for LSMO, including laser and substrate stage scans. Deposition rate was evaluated using RHEED oscillations. A 3 uc thick $LSMO_x$ layer ($0 \leq x \leq 1$) was deposited onto $TiO_2$-terminated STO substrate, followed by a LSMO wedge with thickness variation direction perpendicular to $LSMO_x$ composition gradient. A schematic representation of this sample is represented **Figure 8a**.

M(H) cycles were acquired versus position (512 sites) and temperature (120 temperatures with 90 K < T < 340 K) automatically during a few days. Then $M_{sat}$ and $M_r$ were extracted for each loop, M(T) curves reconstructed and $T_C$ estimated for each (composition x, thickness t) doublet. **Figure 8b** presents the $T_C$ curves plotted versus $t_{LSMO}$ for various Sr content x. One can see that the variation of Tc versus $t_{LSMO}$ depends on x, and in particular for $t_{LSMO} > 7$ uc, the less Sr the more rapid is the $T_C$ decrease. Going from LMO to SMO at $t_{LSMO} = 10$ uc, $T_C$ is increased by 60 K (blue arrows **Figure 8b**). Furthermore, to reach a given Curie temperature of 240 K, one needs 8 uc of LSMO on top of SMO and more than 11 uc of LSMO on top of LMO (red arrows in **Figure 8b**). One can compare these results to the one obtained for x = 0.29 (pink curve in **Figure 8**) where the heterostructure is similar to a simple LSMO/STO interface. Inserting a 3 uc SMO layer at the LSMO/STO interface proves to be beneficial in terms of Tc for $t_{LSMO}$ > 7 uc. However we observe a cross-over for $t_{LSMO} \leq 7$ uc. The Tc decrease with $t_{LSMO}$ accelerates for Sr rich compositions, and no magnetism could be detected at $t_{LSMO}$ = 5 uc for $0.29 \leq x \leq 1$. On the contrary, the lower the Sr content the higher the $T_C$ for $0 \leq x \leq 0.21$ at $t_{LSMO}$ = 5 uc. This reinforcement of FM for LMO coincides with an important increase of the coercive field to values higher than usually observed for LSMO (Hc > 300 Oe at T = 100 K). This is compatible with a second FM phase, harder than LSMO and in contact with it. LMO is antiferromagnetic (AFM) in bulk form. However, several studies reported FM LMO films on STO substrate down to 6 uc. (e.g. [42]) In this article, the transition from AFM to FM has been attributed to an electronic reconstruction at the interface originating from the polar nature of the LMO. In our case the LMO layer is topped by LSMO, and it is quite possible that by proximity effect and/or stress LMO becomes FM at 3 uc thick.

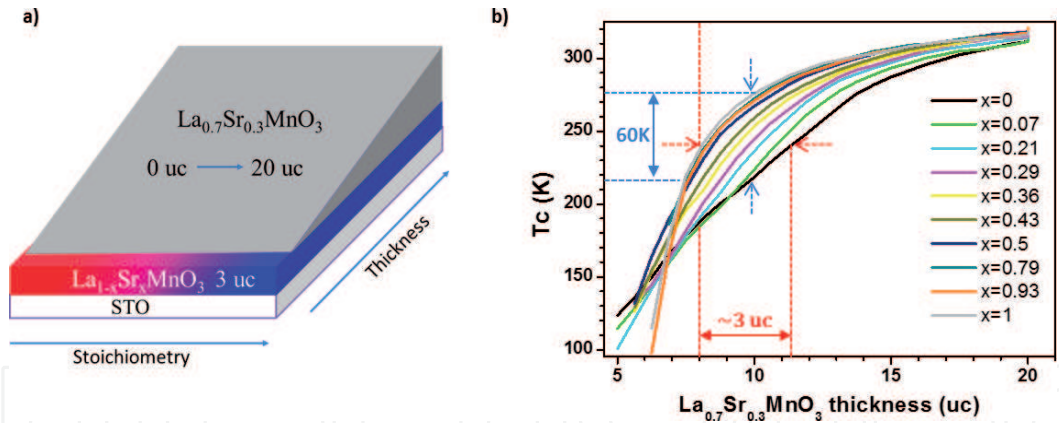

**Figure 8.**
*(a) Schematic representation of the ICPLD $LSMO_x$ / wedge LSMO sample. (b) Curie temperature curves versus $t_{LSMO}$ for various Sr content x in $LSMO_x$ layer.*

*4.2.2 Band alignment at LSMO/STO interface*

We now turn to interface issues arising in tunable capacitors with thinned FE film i.e. the increased influence of dead ferroelectric layer on tunability and the increased leakage current. As discussed previously the insertion of a LSMOx ICPLD layer at LSMO/BST interface may increase interface polarizability and modulate SBH. In order to easily disentangle spontaneous and chemically induced polarizations we choose to work with a non-polar composition of BST i.e. STO. We deposited onto $TiO_2$-terminated STO substrate 38 uc of LSMO followed by 3 uc of $LSMO_x$ ($0 \leq x \leq 1$) and in the direction perpendicular to the gradient a STO wedge (3-15 uc) keeping an access to both LSMO and $LSMO_x$ with the deposition parameters described above.

A schematic of the sample structure is represented **Figure 9a**. The sample was transferred into an ultra-high vacuum atomic force microscope chamber (UHV-AFM Omicron) without breaking the vacuum. The AFM image presented in **Figure 9b** was taken about the red dot in **Figure 9a** with a total thickness of 56 uc. Terraces separated by steps of about 4 Å, i.e. one perovskite cell parameter, are clearly visible (see profile in **Figure 9c**) attesting of the layer by layer growth up to 56 uc. There exists however some 2 Å height features on the terraces indicating the probable existence of two terminations at the surface (SrO and $TiO_2$).

The sample was then air-exposed and inserted into a UV photoelectron spectroscopy chamber (UPS ESCALAB 250Xi Thermo Fisher) to evaluate the work function $W_F$ as a function of position. Although the surface contamination due to air exposure prevented to extract absolute $W_F$ values, the relative variations of $W_F$ with Sr content and STO thickness could be determined assuming a "uniform" surface contamination. UPS spectra were taken at various x content for STO thickness ranging from 0 to 9 uc. A zoom around the emission threshold of the He II UPS spectra (He II energy 40.8 V, bias 4 V) is shown in **Figure 10a** for LSMOx ($t_{STO}$ = 0 uc). From the thresholds one can estimated the $W_F$ reported in **Figure 10b**. A clear continuous decrease of the work function is observed as the Sr content increases. This trend is opposite to the downward Fermi level shift inferred from core-level XPS shift as a function of x reported in the literature [43] and seen by us (not shown).

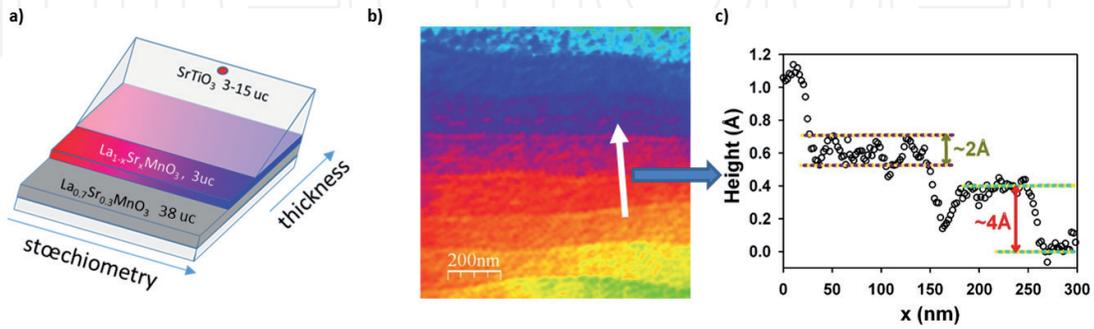

**Figure 9.**
*(a) Schematic of the STO (sub.)/ LSMO / LSMOx / STO heterostructure. (b) UHV-AFM image. (c) Profile from the white arrow in b).*

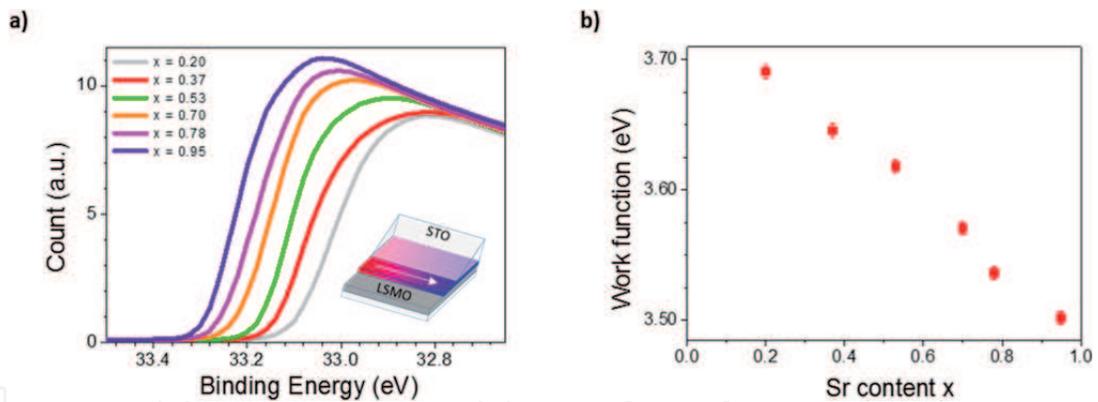

**Figure 10.**
*(a) HeII UPS threshold for various x of air-exposed $LSMO_X$ $U_{bias}$ = 4 V. (b) Corresponding extracted $W_F$ versus Sr content x.*

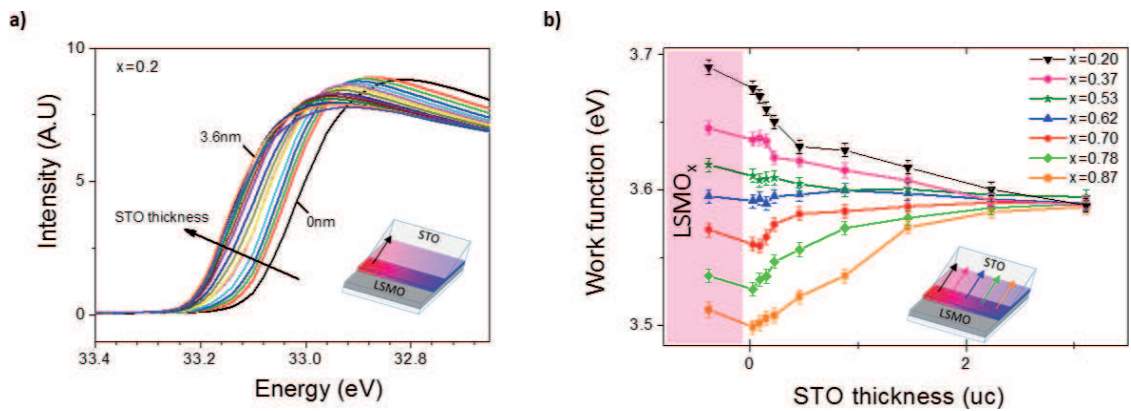

**Figure 11.**
*(a) HeII UPS threshold for various $t_{STO}$ at x = 0.2 $U_{bias}$ = 4 V. (b) Corresponding extracted $W_F$ versus $t_{STO}$ for various Sr content x.*

The counter-intuitive decrease of $W_F$ while $E_F$ decreases too is due to the LSMOx induced charge discontinuity variation at the surface. Going from LMO to SMO, the LSMOx terminal plane changes from $Mn^{3+}O_2^{2-}$ to $Mn^{4+}O_2^{2-}$, i.e. with a surface charge per unit cell going from −1 to 0. The more negatively charged a surface is, the harder for an electron to escape from the surface, the higher the $W_F$. [44].

The electrical nature of the contact between a metal and a semiconductor directly depends on the relative values of the metal $W_F$ and semiconductor electronic affinity $E_a$ for ionic semiconductor [45]. For $E_a > W_F$ an ohmic contact forms, while for $E_a < W_F$ a Schottky barrier is created. STO is generally considered an n-type ionic semiconductor with a Fermi level very close to the conduction band (i.e. $E_a \sim W_F$). As the LSMO$_x$ $W_F$ varies the LSMO$_x$/STO contact nature might be affected. UPS spectra were acquired for various LSMO$_x$ Sr content every 200 μm along the STO wedge. A zoom of the corresponding UPS emission thresholds obtained for x = 0.2 and $0 \leq t_{STO} \leq 9$ uc is shown in **Figure 11a**. The threshold position varies rapidly with $t_{STO}$ for thin STO layers then stabilizes. The $W_F$ estimated from the UPS thresholds for various (x, $t_{STO}$) doublet are reported in **Figure 11b**. One can see the curves folding together toward a $W_F$ value of about 3.58 eV (relative) corresponding to the intrinsic STO work function.

Interestingly this value is inside the range of $W_F$ spanned within LSMO$_x$.(see pink part of **Figure 11b**). Looking at the evolution of $W_F$ vs. $t_{STO}$ for thicknesses up to 3 uc, there is a clear transition from a downward to an upward bending as x increases. This reflects the band bending that occurs at the LSMO$_x$ / STO interface and implies that the contact is modified from Ohmic type to Schottky type.

This result is of importance regarding the optimization of the SBH in BST FE tunable capacitor in particular, but more generally for any metal/semiconductor contacts.

## 5. Conclusion

In this chapter we reviewed the qualities and limitations of PLD for the synthesis of oxides in general and for its use in combinatorial PLD synthesis (CPLD) in particular. We listed some counter-actions to mitigate the PLD limitations together with the mandatory steps to take before attempting reliable CPLD synthesis, i.e. demonstrating the control of both thickness and composition over the whole sample surface. We then detailed a statistical characterization approach to reliably interpret results from CPLD libraries of compounds. An example of this approach is presented, regarding the exploration of lead-free Ga-doped BiFeO$_3$ solid solution for MPB-related piezoelectric properties enhancement. Finally we described a new interface CPLD development (ICPLD) for the exploration of functional interface libraries. This combinatorial interface synthesis approach, with continuous lateral chemical modulation of a few atomic layers, is unique to the best of our knowledge. The effectiveness of ICPLD regarding the control of interface magnetism for magnetic tunnel junctions and energy band and Schottky barrier height tuning in ferroelectric tunable capacitors was demonstrated. This shows that ICPLD is a powerful tool to accelerate heterostructures functional properties enhancement.

## Acknowledgements

Authors would like to thanks J-L Longuet from CEA Le Ripault (France) for the WDS characterizations presented here and Xavier Wallart from IEMN, UMR CNRS 8520, Villeneuve d'Ascq (France), together with Pascal Andreazza from ICMN, UMR CNRS 7374, for UPS measurements.

This work was funded by Région Centre within the projects INTIM-C 2015 and FLEXIGEN.

## Conflict of interest

No conflict of interest.

## Author details

Jérôme Wolfman*, Beatrice Negulescu, Antoine Ruyter, Ndioba Niang and Nazir Jaber†
GREMAN Laboratory, University of Tours, Tours, France

*Address all correspondence to: wolfman@univ-tours.fr

† Current address: LCM Laboratory, CEA Leti, Grenoble, France.

IntechOpen